\documentclass[twocolumn,showpacs,preprintnumbers,nofootinbib]{revtex4}

\usepackage{graphicx}
\usepackage{dcolumn}
\usepackage{bm}
\usepackage{amssymb}  
\usepackage{amsmath}
\usepackage{epsfig}    
\usepackage{here}

\def\beq{\begin{equation}}
\def\eeq{\end{equation}}
\def\bea{\begin{array}}
\def\eea{\end{array}}
\def\be{\begin{equation}}
\def\ee{\end{equation}}
\def\ba{\begin{eqnarray}}
\def\ea{\end{eqnarray}}

\def\to{\rightarrow}

\def\[{\left[}
\def\]{\right]}
\def\({\left(}
\def\){\right)}

\def\mutau{{$\mu N\rightarrow\tau X$~}}
\def\mumu{{$\mu N\rightarrow\mu X$~}}
\def\etau{{$e N\rightarrow\tau X$~}}

\def\sm0{{\widetilde{m}_0}}

\def\U1em{{U(1)_{\rm em}}}
\def\to{\rightarrow}

\def\sq2{\sqrt{2}}

\def\ee{e^+e^-}

\def\End{\end{document}}

\def\Journal#1#2#3#4{{#1} {\bf #2}, #3 (#4)}

\def\PLB{{\rm Phys. Lett.}  B}

\def\PRL{\rm Phys. Rev. Lett.}
\def\PRD{{\rm Phys. Rev.} D}

\def\EPC{{\rm Eur. Phys. J.} C}

\def\MPLA{{\rm Mod. Phys. Lett.} A}

\def\fsl#1{\setbox0=\hbox{$#1$}                 
   \dimen0=\wd0                                 
   \setbox1=\hbox{/} \dimen1=\wd1               
   \ifdim\dimen0>\dimen1                        
      \rlap{\hbox to \dimen0{\hfil/\hfil}}      
      #1                                        
   \else                                        
      \rlap{\hbox to \dimen1{\hfil$#1$\hfil}}   
      /                                         
   \fi}

\allowdisplaybreaks[4]
 
\begin{document}                                                              

\title{A Study of Lepton Flavor Violating 
$\mu N (e N) \to \tau X$ Reactions\\ 
in Supersymmetric Models}%
\author{%
{\sc Shinya Kanemura\,$^1$, Yoshitaka Kuno\,$^1$,
     Masahiro Kuze\,$^2$, Toshihiko Ota\,$^1$}
}
\affiliation{%
\vspace*{2mm} 
$^1$Department of Physics, 
Osaka University, Toyonaka, Osaka 560-0043, JAPAN\\
$^2$Department of Physics, Tokyo Institute of Technology, \\
Tokyo 152-8851, JAPAN\\
}

\begin{abstract}
\hspace*{-0.35cm}
We study a lepton flavor violating \mutau reaction in deep 
inelastic scattering region in supersymmetric models. 
The contribution from the Higgs boson mediation could be 
important for this reaction. For that case, the cross section 
is constrained by the experimental limit of the pseudo-scalar 
coupling from $\tau\rightarrow \mu\eta$ decays. 
We find that at a muon energy ($E_{\mu}$) higher than 50 GeV, 
the predicted cross section increases significantly due to 
the contribution from sea $b$-quarks.
As a result, with $10^{20}$ muons per year, at most  
a number of $\mathcal{O}(10^4)$ is expected for \mutau events
at $E_{\mu}$= 300 GeV, whereas 
$\mathcal{O}(10^2)$ events are given at $E_{\mu}= 50$ GeV. 
Furthermore, the \mutau phenomenology, 
in particular that for the signal and 
backgrounds, is briefly discussed. 
Another promising possibility to search for the \etau reaction 
at an electron-positron linear collider is also discussed. 
Searches for these reactions would be competitive to 
studies of rare tau decays and have potential to improve 
sensitivities to lepton flavor violation significantly.
\pacs{11.30.Hv; 12.60.Jv; 14.60.Fg. 
\hfill   ~~ [ \today ~and~ hep-ph/0410044 ] }
\end{abstract}

\maketitle

\setcounter{footnote}{0}
\renewcommand{\thefootnote}{\arabic{footnote}}

\section{Introduction} 
Lepton flavor violation (LFV) for charged leptons 
is clear evidence to indicate physics beyond the standard model.
A type of models based on supersymmetry (SUSY) naturally induces 
LFV couplings at loop levels due to the slepton mixing~\cite{bm,Tohoku}. 
The predicted branching ratios of LFV processes in SUSY models
are only a few orders of magnitude
smaller than the present experimental limits, and therefore 
they could be reached by future experiments. 
Various
LFV processes can be experimentally searched. 
For the electron-muon sector, rare muon decay processes 
such as $\mu \to e \gamma$ and $\mu \to eee$ as well as 
$\mu-e$ conversion in a muonic atom have been studied and will be 
further examined by forthcoming experiments~\cite{muonLFV}.

In SUSY models, there are two types of 
LFV couplings: namely
those mediated by the neutral gauge bosons
and those by the neutral Higgs bosons. In particular, 
the Higgs-mediated
LFV couplings attract much interest recently, because their contributions
do not decouple even if the soft SUSY breaking scale is 
as large as $\mathcal{O}(1)$ TeV~\cite{Babu,Ellis,Rossi,mu-e-conv}. 
It is in contrast to 
the case of the gauge boson mediation.
Since the Higgs-mediated LFV couplings are proportional 
to the mass of the relevant charged leptons, 
the tau-associated processes are useful to study them.
In recent years, the LFV couplings associated with a tau lepton 
have been measured at $B$ 
factories~\cite{Belle-Tau-MuEta,Belle-Tau-MuGamma,Belle-Tau-3Mu} 
by searches for rare tau decays, such as  
$\tau \to \mu \gamma$, 
$\tau \to 3 \mu$, 
$\tau \to \mu \pi\pi$, 
$\tau \to \mu \eta$, etc.
Direct searches for the Higgs LFV coupling via decays of the Higgs
bosons ($\Phi^{0}$),
$\Phi^{0} \rightarrow \tau^{\pm} \mu^{\mp}$, 
have been proposed by several authors, either
at CERN Large Hadron Collider (LHC)~\cite{Assamagan} 
or at a Linear Collider (LC)~\cite{Higgs-TauMu-LC}. 

In this letter, 
we would like to discuss an alternative approach of searching 
for the LFV couplings associated with a tau lepton
by a \mutau reaction at the deep inelastic scattering (DIS) region  
with high-intensity and high-energy muon beams. 
Here, $N$ is a target nucleon, and $X$ represents 
all final state particles. Recently, Sher and Turan have 
discussed this process in a model 
independent approach~\cite{Sher-DIS}.  
Instead, we here study this process in the 
framework of SUSY, and evaluate the cross section 
under the current experimental constraints from the other LFV limits.

The $\mu-\tau$ transition processes have been estimated 
by the effective LFV couplings.
The upper limits on the four-Fermi LFV couplings
have been studied without assuming specific theoretical models 
in Ref.~\cite{Black-Han-He-Sher}.
When the scalar LFV coupling is independent of the other types of
couplings, its experimental constraint 
only comes from the process in which a tau decays into a muon and 
a pion pair ($\tau \to \mu\pi\pi$). 
The total cross section of the process \mutau mediated by the 
scalar LFV coupling could then be as large as 0.5 fb at muon energy 
($E_{\mu}$) of 50 GeV~\cite{Sher-DIS}. 
For this case, with $10^{20}$ muons per year on a 
$\mathcal{O}(10^2)$ g/cm$^2$ target mass, about 
$10^{6}$ of the \mutau events can be produced, or no observation of the 
\mutau signal would improve the limits by six orders of 
magnitude~\cite{Sher-DIS}.
However, in SUSY models, the effective scalar coupling 
which is mediated by the CP-even Higgs bosons ($h^{0}$ and $H^{0}$) 
is correlated with the pseudo-scalar coupling 
which is mediated by the CP-odd Higgs boson ($A^{0}$). 
Therefore, the constraint on the pseudo-scalar coupling from 
the $\tau \to \mu \eta$ result~\cite{Belle-Tau-MuEta} 
must be applied to the scalar coupling~\cite{Sher-Tau-MuEta}.
Consequently, the predicted \mutau cross section 
would become smaller than that in Ref.~\cite{Sher-DIS}.

The \mutau cross section is known to increase as $E_{\mu}$ becomes high. 
When $E_{\mu}$ is high enough, for instance more than several GeV, 
the DIS reaction dominates. 
Therefore, we calculate the \mutau cross section at the DIS region
by using the realistic parton distribution function (PDF), 
and in particular estimate the contributions from different 
quarks separately. 
In SUSY models, the contributions from the down-type quarks are 
important for the Higgs-mediated LFV interaction for the region 
of large $\tan\beta$, where $\tan\beta$ is the ratio of the Higgs 
vacuum expectation values. 
Since the Yukawa couplings of the Higgs bosons with quarks 
are proportional to the mass of the associated quarks,
the contribution of the $d$-quark is not large, although
its PDF contribution is large.  
For $E_{\mu}$ of a few GeV, the contribuion 
from the $s$-quark sub-process, $\mu  s \to \tau  s$, dominates.
We find that at energies higher than about 50 GeV, the contribution from 
the $b$-quarks arising from the gluon splitting becomes important. 
Therefore, at $E_\mu =100$ GeV
the cross section of \mutau can be one order of magnitude greater than 
that at $E_\mu =50$ GeV.

This paper is organized as follows. 
In the following section, we describe the LFV interaction in
the minimum supersymmetric standard model (MSSM), 
and calculate the maximally allowed \mutau cross sections
at the DIS region with experimental constraints.  
In the subsequent section, the \mutau phenomenology, in particular that 
for the signal and background, is discussed. 
In Sec.~IV, the \etau search at a LC to study the $e-\tau$ transition 
is proposed. Conclusions are presented in Sec.~V.

\section{ The \mutau Process in Minimum Supersymmetric Standard Model}
The effective Lagrangian describing the sub-process 
$\mu q \rightarrow \tau q$, where $q$ is a quark,  
in the MSSM is given by 
\begin{align}
\mathcal{L}_{\text{eff}}
=&
\sum_{q} \sum_{X} 
\left\{
(\mathcal{A}_{X}^{V})_{q}
 \left(
  \overline{\tau} \gamma^{\rho} {\rm P}_{X}^{} \mu 
 \right)
 \left(
  \bar{q} \gamma_{\rho} q
 \right)\right. \nonumber \\
& +
(\mathcal{A}_{X}^{T})_{q}
 \left(
 \overline{\tau} \frac{{\rm i} m_{\tau} q_{\sigma}
  \sigma^{\rho\sigma}}{q^{2}} {\rm P}_{X}^{} \mu 
 \right)
 \left(
  \bar{q} \gamma_{\rho} q
 \right) \nonumber \\
&+
(\mathcal{C}_X^{hH})_{q}
 \left(\overline{\tau} {\rm P}_{X}^{} \mu \right)
 \left(\overline{q} q\right) \nonumber \\
& \left.+
 (\mathcal{C}_X^{A})_{q}
 \left(\overline{\tau} {\rm P}_{X}^{} \mu \right)
 \left(\overline{q}\gamma^{5} q\right)\right\},  \label{eq1}
\end{align}
where $\mathcal{A}_{X}^{T,V}$ are the effective couplings for 
the gauge boson mediation with  
the projection operator ${\rm P}_X$ 
for the chirality $X=L$ and $R$.    
The scalar coupling $\mathcal{C}^{hH}_X$ and the pseudo-scalar 
coupling $\mathcal{C}^{A}_X$ are those mediated 
by $h^0$ and $H^0$, and by $A^0$, respectively. 
They are induced due to quantum corrections in the MSSM. 
The details can be found in Ref.~\cite{Tohoku} 
for the gauge boson mediation, 
and in Ref.~\cite{mu-e-conv} for the Higgs boson mediation. 
In the following, the cross section of the 
\mutau reaction is calculated separately for the cases of 
the Higgs boson mediation and the gauge boson mediation,  
since the SUSY parameter regions where each of them is enhanced
are different. 

First, let us start with the Higgs boson mediation.
It is known that this type of effect does not decouple 
even in the large $m_{\text{SUSY}}$ limit, where $m_{\text{SUSY}}$ is a
typical scale of the soft SUSY breaking~\cite{Babu,Ellis,Rossi}. 
The contribution of the Higgs boson mediation to the differential cross
section $\mu^- N \rightarrow \tau^-  X$ is given by
\begin{align}
\frac{{\rm d}^{2} \sigma}{{\rm d} x {\rm d} y}
& =
  \sum_{q} x f_{q} (x) 
 \left\{
 \left( 
 \left|{\mathcal{C}^{hH}_L}\right|^{2} + 
 \left|{\mathcal{C}^{A}_L} \right|^{2} \right)_{q} 
 \left(
 \frac{1-\mathcal{P}_\mu}{2}
\right) 
\right. \nonumber\\
&+
\left. 
 \left( 
 \left|{\mathcal{C}^{hH}_R}\right|^{2} + 
 \left|{\mathcal{C}^{A}_R} \right|^{2} \right)_{q} 
 \left(
 \frac{1+\mathcal{P}_\mu}{2}
\right) 
\right\}
 \frac{s}{16 \pi} 
  y^{2}, \label{cross-higgs}
\end{align}
where the function $f_{q}(x)$ is the PDF 
for $q$-quarks,
$\mathcal{P}_\mu$ is the incident muon polarization such that  
$\mathcal{P}_\mu=+1$ and $-1$ correspond to the 
right- and left-handed polarization, respectively, 
and $s$ is the center-of-mass (CM) energy. 
The parameters $x$ and $y$ are defined as
\begin{align}
x\equiv \frac{Q^{2}}{2 P \cdot q}, \qquad 
y\equiv \frac{2 P \cdot q}{s}, 
\end{align} 
in the limit of massless tau leptons, where $P$ 
is the four momentum of the target, $q$ is the momentum
transfer, and $Q$ is defined as $Q^{2} \equiv -q^{2}$.
As seen in Eq.~(\ref{cross-higgs}), experimentally,   
the form factors of 
${\mathcal{C}^{hH}_L}$ and ${\mathcal{C}^{A}_L}$
(${\mathcal{C}^{hH}_R}$ and ${\mathcal{C}^{A}_R}$)  
can be selectively studied by using purely left-handed
(right-handed) incident muons.
In SUSY models such as the MSSM with heavy right-handed 
neutrinos, LFV is radiatively induced due to the left-handed 
slepton mixing, which only affects $\mathcal{C}_L^{hH}$ 
and $\mathcal{C}_L^{A}$.  
Therefore, in the following, 
we focus only on those $\mathcal{C}_L^{hH}$ 
and $\mathcal{C}_L^{A}$ couplings\footnote{
It is known that 
in some models, such as $SU(5)$ grand unified theories, 
$\mathcal{C}_R^{hH}$ and $\mathcal{C}_R^{A}$  
can be induced due to the right-handed slepton
mixing~\cite{hisano-nomura}.
}.  

The magnitudes of the effective couplings are constrained by the current 
experimental results of searches for LFV processes of tau decays. 
In the case that all the effective couplings are independent, the 
constraints for the vector, axial-vector, scalar, and pseudo-scalar
couplings are given in Ref.~\cite{Black-Han-He-Sher}. 
In the MSSM, however, the couplings are related each other. In particular, 
in the decoupling region ($m_{A} \gtrsim 150$ GeV),
the scalar coupling $\mathcal{C}^{hH}_L$ 
is nearly equal to the pseudo-scalar coupling $\mathcal{C}^{A}_L$, since 
$m_{H} \simeq m_{A}$ and $\sin(\alpha-\beta) \simeq -1$,
where $\alpha$ is the mixing angle for the CP-even Higgs bosons.
Therefore, both couplings are determined by the one that is 
more constrained, namely the pseudo-scalar coupling. 
It is constrained by the
$\tau \rightarrow \mu \eta$ decay  
(${\rm Br}(\tau \to \mu \eta) 
< 3.4 \times 10^{-7}$)~\cite{Belle-Tau-MuEta}. 
Then the constraint is given on the $s$-associated 
scalar and pseudo-scalar couplings in Eq.~(\ref{eq1}) by
\begin{align}
  \left(\left|{\mathcal{C}^{hH}_L}\right|^{2} 
+ \left|{\mathcal{C}^{A}_L}\right|^{2}\right)_{s}
 &\simeq
2 (\left|{\mathcal{C}^{A}_L}\right|^{2})_{s} \nonumber \\
&\hspace{-1.7cm}\lesssim 1.9 \times 10^{-9} [\text{GeV}^{-4}] \times
 \text{Br}(\tau \rightarrow \mu  \eta).
\label{eq:exp-limit-on-CA}
\end{align} 
The largest values of ${\mathcal{C}^{hH}_L}$ 
and ${\mathcal{C}^{A}_L}$
can be realized with
$m_{\text{SUSY}} \sim \mathcal{O}(1)$ TeV and 
the Higgsino mass $\mu \sim \mathcal{O}(10)$ TeV~\cite{Rossi}.
It should be noted that in such a situation, the gauge boson mediated
couplings are strongly suppressed. 

\begin{figure}
\vspace{0.5cm}
\includegraphics[width=8cm]{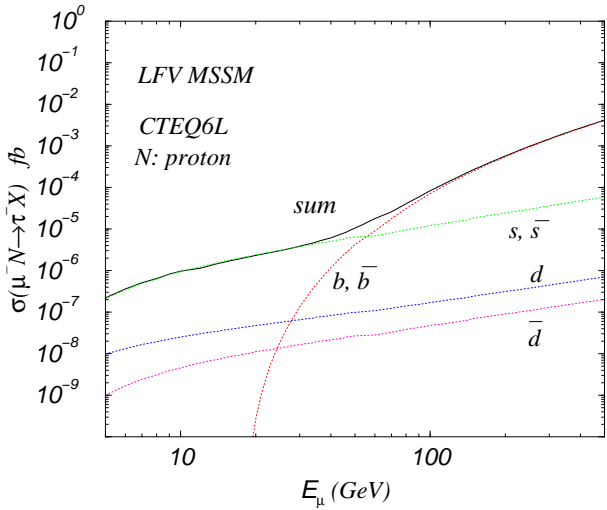}
\caption{%
Cross section of the $\mu^- N \to \tau^- X$ DIS process 
as a function of the muon energy for 
the Higgs mediated interaction. It is assumed that the initial 
muons are purely left-handed.
CTEQ6L is used for the PDF. 
}\label{Fig:total-cross-section-vs-Emu}
\end{figure}

We evaluate the cross sections of 
the \mutau reaction in the DIS region for the Higgs-mediated interaction 
with the maximally allowed values of the effective couplings 
in Eq.\eqref{eq:exp-limit-on-CA}  as a reference. 
They are  plotted in Fig.\ref{Fig:total-cross-section-vs-Emu} 
for different quark contributions as a function of the 
muon beam energy in the laboratory frame. For the PDF, 
we used CTEQ6L~\cite{CTEQ6} in our analysis. 
The target $N$ is assumed to be a proton. 
For a nucleus target, the cross section would be higher,  
approximately by the number of nucleons in the target.

The cross section sharply increases above $E_{\mu} \sim 50$ GeV
in Fig.\ref{Fig:total-cross-section-vs-Emu}.
This enhancement comes from a consequence of the $b$-quark
contribution in addition to the $d$ and $s$-quark contributions. 
The coupling for the $b$-quark is enhanced by a factor of
$m_{b}/m_{s}$ over the $s$-quark contribution as given by
\begin{align}
(\left|{\mathcal{C}^{A}_L}\right|^{2})_{b} = 
 \left(
 \frac{m_{b}}{m_{s}}
 \right)^{2}
 (\left| {\mathcal{C}^{A}_L} \right|^{2})_{s}.
\end{align}
The cross section is enhanced by one order of magnitude when the muon
energy changes from 50 GeV to 100 GeV. Typically, 
for $E_{\mu} = 100$ GeV and $E_{\mu} = 300$ GeV, the cross section 
is $10^{-4}$ fb and $10^{-3}$ fb, respectively. 
In our analysis, we used
$m_b=4.3$ GeV and $m_s=120$ MeV.

Next we study the case where the gauge-boson mediated
interaction is dominant, for instance, that with   
$m_{\text{SUSY}} \sim \mathcal{O}(100)$ GeV~\cite{Ellis}. 
The differential cross sections for \mutau with 
the tensor couplings $\mathcal{A}_{L,R}^T$ and 
the vector couplings $\mathcal{A}_{L,R}^V$ are calculated as  
\begin{align}
\left.\frac{{\rm d}^{2} \sigma}{{\rm d} x {\rm d} y}\right|_{\rm tensor}
&= 
\sum_{q} x f_{q} (x) \left\{
 \left(
 |\mathcal{A}^{T}_{R}|^{2} + |\mathcal{A}^{T}_{L}|^{2}
 \right)  
\right. \nonumber\\
&
\left.
 + \mathcal{P}_\mu
 \left(
 |\mathcal{A}^{T}_{R}|^{2} - |\mathcal{A}^{T}_{L}|^{2}
 \right) 
\right\}_{q}
 \frac{m_{\tau}^{2} }{8 \pi} \frac{1}{xy} (1-y), \\
\left.\frac{{\rm d}^{2} \sigma}{{\rm d} x {\rm d} y}\right|_{\rm vector}
&= 
\sum_{q} x f_{q} (x) \left\{
 \left(
 |\mathcal{A}^{V}_{R}|^{2} + |\mathcal{A}^{V}_{L}|^{2}
 \right) 
\right. \nonumber\\
&
\left.
 + \mathcal{P}_\mu
 \left(
 |\mathcal{A}^{V}_{R}|^{2} - |\mathcal{A}^{V}_{L}|^{2}
 \right) 
\right\}_{q}
 \frac{s}{16 \pi} \left\{1 + (1-y)^2\right\}, 
\end{align}
respectively. 
The effective tensor couplings are strongly constrained by 
the $\tau \to \mu \gamma$ process~\cite{Belle-Tau-MuGamma}, as 
\begin{gather}
\left(
 \left| \mathcal{A}_{R}^{T} \right|^{2}
 +
 \left| \mathcal{A}_{L}^{T} \right|^{2}
\right)_{d,s,b}
 \lesssim 
6.4 \times 10^{-14} [\text{GeV}^{-4}]
\times \text{Br}(\tau \rightarrow \mu\gamma) 
\end{gather}
and 
\begin{gather}
\left(
 \left| \mathcal{A}_{R}^{T} \right|^{2}
 +
 \left| \mathcal{A}_{L}^{T} \right|^{2}
\right)_{u,c}
= 4 \times \left(
 \left| \mathcal{A}_{R}^{T} \right|^{2}
 +
 \left| \mathcal{A}_{L}^{T} \right|^{2}
\right)_{d,s,b}.
\end{gather}
Since Br$(\tau \rightarrow \mu\gamma)< 3.1\times 10^{-7}$ 
\cite{Belle-Tau-MuGamma}, 
the contribution from the tensor interaction is found to be smaller 
than that from the Higgs boson mediation by about five orders of magnitude. 
On the other hand, the vector and axial-vector interactions 
are suppressed at the same level as the pseudo-scalar 
interaction~\cite{Black-Han-He-Sher}.
Therefore, 
their contributions can be as large as those for the Higgs boson mediation, 
if $E_\mu$ is less than than 50 GeV~\cite{Sher-DIS}.   
For instance, the cross section from the vector (or axial vector) 
interaction can be of the order of $10^{-4}$ fb for $E_\mu=50$ GeV.
At higher energies, the cross section for 
the gauge boson mediation are much smaller than 
those for the Higgs boson mediation because of no enhancement 
by the $b$-quark sub-process.

It is concluded that the DIS process \mutau can be more useful 
to search the Higgs mediated LFV interaction in the MSSM
for higher energy muon beams.

\section{The \mutau Phenomenology} 

With the intensity of $10^{20}$ muons per year and 
the target mass of 100 g/cm$^2$, 
about $10^4$ ($10^2$) events could be 
expected for $\sigma(\mu N\rightarrow \tau X)=10^{-3} ~(10^{-5})$ fb, 
which corresponds to $E_{\mu}=300$ $(50)$ GeV 
from Fig.~\ref{Fig:total-cross-section-vs-Emu}. 
This would provide good potential to improve the sensitivity by four (two)
orders of magnitude from the present limit from 
$\tau\rightarrow\mu\eta$ decay, 
respectively. Such a muon intensity could be available at a future
muon collider (MC)~\cite{MC} and a neutrino factory (NF)~\cite{NF}. 

\begin{figure*}
\includegraphics[width=8cm]{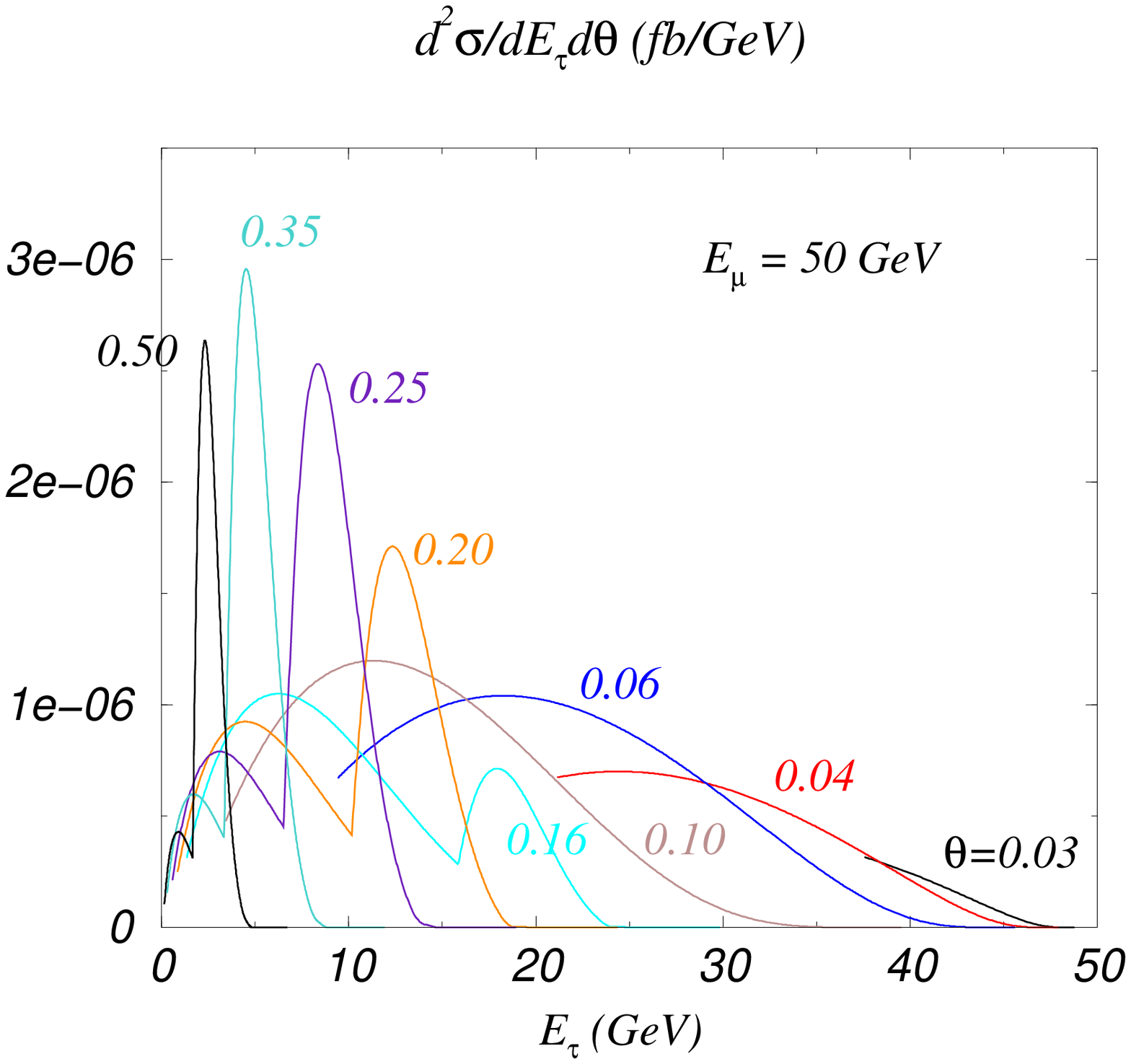}
\hspace{5mm}
\includegraphics[width=8cm]{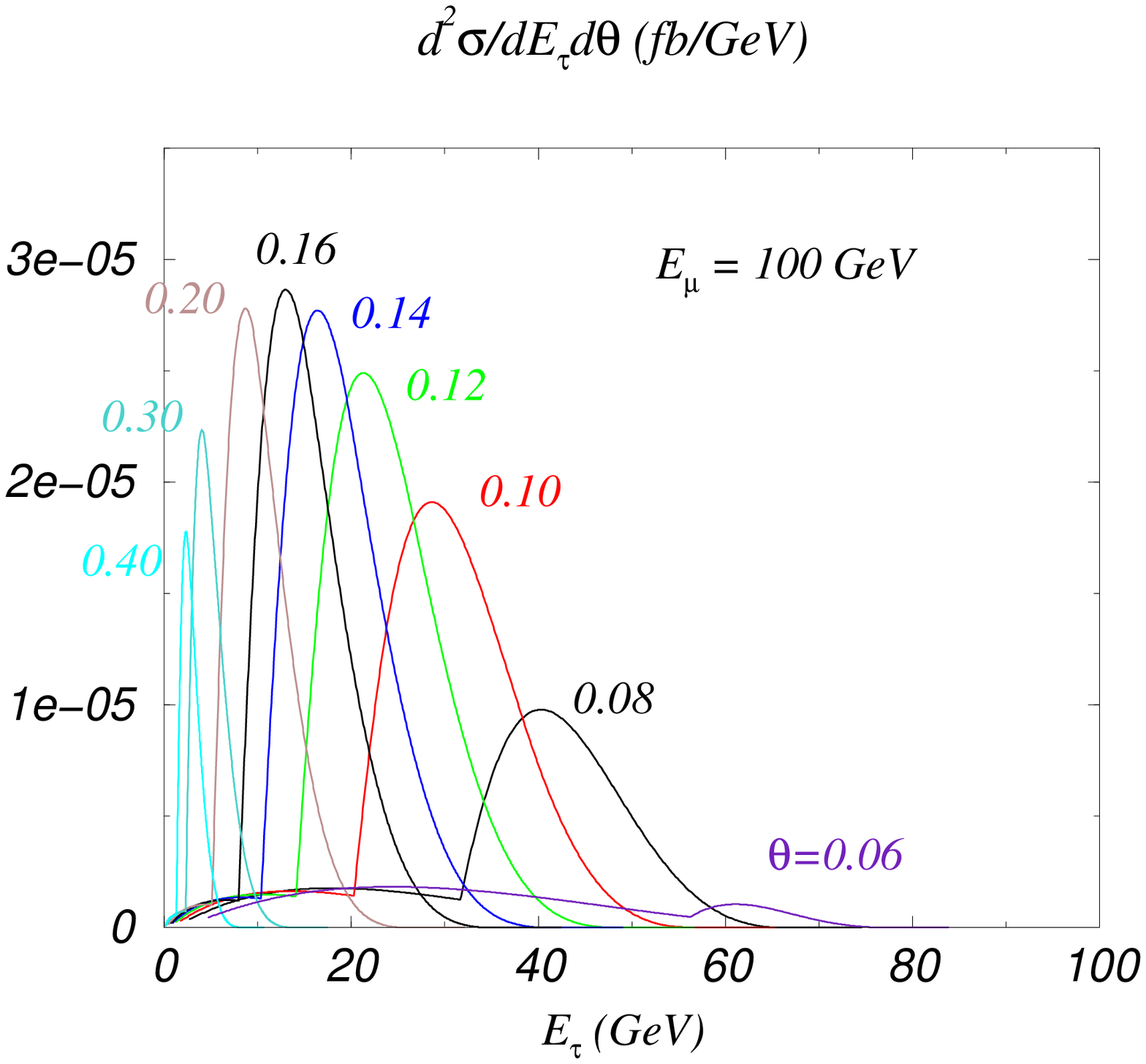}
\caption{%
The differential cross section of the tau 
from the $\mu^- N \to \tau^- X$ DIS process as a function 
of the tau energy ($E_{\tau}$) and the tau emission angle 
($\theta$) with respect to the forward direction 
for $E_{\mu} = 50$ GeV (left) and $E_{\mu} = 100$ GeV (right).
It is assumed that the initial muons are purely left-handed.
}
\label{Fig:angle-dependence}
\end{figure*}

We have studied the signal events of the \mutau reaction.
In the Higgs boson mediated interaction, 
the tau leptons in the \mutau reaction 
are emitted at a relatively large angle with respect to the beam
 direction. It is in contrast to the gauge mediated interaction where 
the tau leptons are forward-peaked.
The differential cross section is 
given as a function of the tau energy ($E_{\tau}$) and the 
tau emitted angle ($\theta$) from the beam direction by 
\begin{align}
\frac{{\rm d}^{2} \sigma}{{\rm d} E_{\tau} {\rm d} \theta} =
\frac{{\rm d}^{2} \sigma}{{\rm d} x {\rm d} y} \cdot
\frac{2 E_{\mu} E_{\tau} \sin\theta}{s (E_{\mu} - E_{\tau})}. 
\end{align}
In Fig.~\ref{Fig:angle-dependence}, the $E_\tau$ dependence 
in the differential cross section is shown for each $\theta$ at 
$E_{\mu}$=50 GeV and $E_{\mu}$=100 GeV. 
For $E_{\mu} = 50$ GeV, there are two peaks 
for each emission angle $\theta$ of the tau leptons. 
The peak at lower $E_\tau$ for each curve 
corresponds to the contribution 
from the $s$-quark sub-process, and the other 
does to the $b$-quark contribution.
The tau leptons from the $b$-quark sub-process
are emitted with larger emission angles with relatively 
low energies, while those from the $s$-quark contribution
carry higher energies with smaller emission angles.
For $E_{\mu} = 100$ GeV, the contribution from the 
$b$-quark sub-process is dominant. The tau leptons are 
emitted with large angles around $\theta=0.16$. 

To identify the tau lepton from the \mutau reaction, direct 
measurement of tau lepton tracks (such as by emulsions) might not be 
possible at such a high beam rate. 
Instead, the identification might be possible 
by tagging the tau decay products and observing their decay kinematics. 
Among various decay modes, one might consider leptonic decays of the 
tau leptons as discussed in Ref.~\cite{Gninenko}. 
Another candidate could be 
to detect a hadron from the two-body tau decays.  
The branching ratios, such as 
$\tau \rightarrow \nu_{\tau} \pi$, $\nu_{\tau} \rho$
and $\nu_{\tau} a_{1}$, are about 0.3 in total. 
In SUSY models with left-handed slepton mixing, 
the $\tau^{-}$($\tau^+$) produced through the Higgs-mediated interaction 
is only right-handed (left-handed) for an incident left-handed 
$\mu^-$ beam (right-handed $\mu^+$ beam).
The hadrons from the right-handed $\tau^{-}$ decay (left-handed $\tau^+$     
decay) tend to be emitted in the direction of the parent tau lepton, 
and therefore be rather energetic~\cite{tau-R-hadron-decay}. 
At the same time, the neutrino from the tau decay would carry 
a part of the energy of the tau lepton, resulting in missing energy.
Therefore, the signature of the events could be a hard hadron 
from decay of the tau lepton produced at 
a relatively large angle from the beam direction (namely a 
hadron with large transverse momentum $p_{T}$) and some missing energy. 
Those hadrons from the tau decay should be discriminated  
from the hadrons from the target nucleons  
which have mostly soft energies.

Discrimination of pions from muons could be the most critical issue, 
since a muon from the DIS process \mumu 
would mimic the signal if it is misidentified 
as a pion. Therefore, they should be rejected with high efficiency.
The rate of the DIS process \mumu is very high, of the order of 
$\mathcal{O}(1)$ $\mu$b in total. 
Since the muon from \mumu is very forward peaked, 
elimination of forward-going particles would give rejection in addition to the
particle identification with a cost of the signal acceptance. The rejection of
muons and the acceptance of hadrons from 
the tau leptons in \mutau might be optimized
in consideration of the aimed signal sensitivity.
In either case, realistic Monte Carlo studies should be 
necessary\footnote{
Studies using a modified version 
of LQGENEP generator~\cite{lqgenep} are underway.}.

It should be noted that in the models where the scalar and 
pseudo-scalar couplings are independent, such as in the 
two Higgs doublet model, the experimental constraints become rather weak, 
in contrast to the MSSM models which we have so far discussed.
The cross sections allowed could become larger by about $\mathcal{O}(10^5)$, 
yielding $\sigma(\mu N \rightarrow \tau X) \sim \mathcal{O}(10)$ fb 
at the muon energy of $\mathcal{O}(100)$ GeV~\cite{Sher-DIS}.
At present, the muon beam line at CERN can provide a muon beam 
with energy up to 190 GeV and $10^{14}$ muons per year~\cite{CERN190GeVmuon}. 
By using this, for instance,
we could expect about $\mathcal{O}(10^2)$ events from 
Fig.\ref{Fig:total-cross-section-vs-Emu},
or improve the effective scalar coupling by about $\mathcal{O}(10^2)$. 

If incident muons are polarized, T-odd correlations between the muon spin 
and the tau momentum  and spin polarization vectors can be examined in the \mutau reaction. 
The T-odd correlation would give opportunity to study CP violation in 
SUSY. They will be studied in our coming publications.

\section{Probing $e-\tau$ transition via $eN \rightarrow \tau X$ reaction}

We would like to comment on another LFV search for $e-\tau$ transition 
through \etau with a 
high energy electron (or positron) beam. 
In the MSSM, similar to the $\bar \tau_R^{} \mu_L^{} \Phi^0$ coupling, 
the $\bar \tau_R^{} e_L^{} \Phi^0$ coupling
is proportional to the tau lepton mass,
where $\Phi^0$ could be either $h^0,H^0$ or $A^0$, 
and $l_{R}$ and $l_{L}$ ($l$=$e$, $\mu$, $\tau$) are the corresponding 
right-handed and left-handed leptons, respectively.
The coupling of $\bar \tau_R^{} e_L^{} \Phi^0$ could be
as large as that of $\bar \tau_R^{} \mu_L^{}  \Phi^0$, 
when the slepton mixing between 
$\tilde{\tau}_L^{}$ and $\tilde{e}_L^{}$ is similar to that between 
$\tilde{\tau}_L^{}$ and $\tilde{\mu}_L^{}$. Therefore, our discussions 
on the \mutau DIS reaction can be directly applied to the case
of \etau reaction.
Experimental constraints on the effective 
form factors for $e-\tau$ transition are given by 
$\tau \to e \eta$ and $\tau \to e \pi\pi$ decays 
at the $B$ factories~\cite{pdg2004} and the
$e-\tau$ searches at HERA~\cite{HERA}. 
The strongest bound on the pseudo-scalar coupling for 
the $e-\tau$ transition comes from the $\tau \to e \eta$ 
result, which is similar to the bound on the $\mu-\tau$ 
transition from the $\tau \to \mu \eta$ result. 
For the scalar coupling, the results of the $e p \to \tau X$ search 
at HERA give stronger bounds than the 
$\tau \to e \pi\pi$ result. 
In the MSSM, therefore, the maximally-allowed cross section for \etau 
is in the same order of magnitude as that for \mutau 
for the same incident beam energy.

At a future electron-positron LC with the 
CM energy of $500$ GeV and the luminosity 
of $10^{34}$ cm$^{-2}$ s$^{-1}$, 
$10^{22}$ electrons (or positrons) with the energy 
250 GeV would be available per year.
Assuming that they could be bombarded on a fixed target with 
the mass 10 g/cm$^2$, 
we could expect about $10^{4-5}$ of \etau events 
for $\sigma(eN\to\tau X) \sim 10^{-3}$ fb. 
Therefore, significant improvement,
by several orders of magnitude, on the $e-\tau$ coupling can be expected
over the present limits.  
Experimental 
issues in the \etau search, 
such as background rejections, would be very different from 
the \mutau case. Further
studies will appear in our future publications.

Combining both the $e-\tau$ searches at a LC 
and the $\mu-\tau$ searches at a MC and a NF, we could 
study the slepton mixing and SUSY breaking in more details. 

\section{Conclusions}

We have calculated the cross sections for \mutau at
the DIS region in the MSSM separately
for the Higgs boson mediation and the gauge boson
mediation. 
We have found that this process can be useful 
to search for the Higgs-boson mediated LFV coupling. 
In the MSSM, the Higgs boson mediation is constrained from 
the $\tau \to \mu\eta$ result, so that the cross section 
is at most $10^{-5}$ fb at the incident muon energy $E_\mu$ of 
$50$ GeV. 
If $E_\mu$ is higher than 50 GeV, 
contributions from the sea $b$-quarks become more significant,  
and thereby the cross section is drastically enhanced. 
For $E_\mu=300$ GeV, for instance, 
the cross section can be as large as $10^{-3}$ fb. 
Experimental issues on the tau identification and background rejection 
have been briefly examined. 
We also studied the \etau transition to search for the possible $e-\tau$
transition at a future LC, and expect significant improvements. 
Once high-intensity and high-energy muon and electron beams are available, 
there would be better opportunity to improve the limit on the 
LFV couplings for the Higgs boson mediation by a few orders of magnitude. 
Searches for the  \mutau and \etau reactions would be competitive to 
studies of  rare tau decays at planned 
future super $B$ factories.
 
The authors would like to thank Prof. Yasuhiro Okada, 
Dr. Lorenzo Bellagamba and Dr. Hiroyuki Kawamura 
for their useful comments and discussions.

\end{document}